\documentclass[aps,prl,reprint]{revtex4-2}

\usepackage{graphicx}% Include figure files
\usepackage{dcolumn}% Align table columns on decimal point
\usepackage{bm}% bold math
\usepackage{xcolor}  % Include the xcolor package for colors

\begin{document}
\flushbottom
\title{Laser-Induced Quenching of the Th-229 Nuclear Clock Isomer in Calcium Fluoride}

\author{F. Schaden}
%\thanks{These authors contributed equally to this work.}
\author{T. Riebner}
\author{I. Morawetz}
\author{L. Toscani De Col}
\author{G.A. Kazakov}
\author{K. Beeks}
\author{T. Sikorsky}
\author{T. Schumm}
\email{email: thorsten.schumm@tuwien.ac.at}
\affiliation{Vienna Center for Quantum
Science and Technology, Atominstitut, TU Wien, 1020 Vienna, Austria}

\author{K. Zhang}
%\thanks{These authors contributed equally to this work.}
\author{V. Lal}
\author{G. Zitzer}
\author{J. Tiedau}
\author{M.V. Okhapkin}
\author{E. Peik}
\email{email: ekkehard.peik@ptb.de}

\affiliation{Physikalisch-Technische Bundesanstalt, 38116 Braunschweig, Germany}

\date{\today}
\begin{abstract}
The 10-minute radiative lifetime of the first excited $^{229}$Th$^{4+}$ nuclear state in ionic crystals provides narrow spectroscopic linewidths, enabling the realization of a solid-state nuclear clock. Due to the 4+ noble gas configuration, electronic readout or state initialization schemes known from atomic clocks are inaccessible. This elongates the interrogation cycle, which will deteriorate the clock performance. To address this limitation we demonstrate laser-induced quenching (LIQ) as a method of depumping the $^{229}$Th isomer population in CaF$_2$.
We provide experimental evidence for LIQ at different wavelengths (148\,--\,420\,nm) and temperatures (100\,--\,350\,K), achieving a threefold reduction in the isomer lifetime with 20\,mW of laser power.
\end{abstract}

\maketitle
The nuclear isomer state of $^{229}$Th is intensively studied due to its unique low excitation energy at approximately 8.4\,eV~\cite{Beeks:2021}. The first laser excitation of this nucleus was recently demonstrated~\cite{Tiedau:2024} and quickly confirmed~\cite{UCLA2024}. Additionally, the lifetime of its excited state $^{229m}$Th has been determined in both vacuum and various solid-state environments~\cite{Kraemer:2022,Yamaguchi:2024,Tiedau:2024,UCLA2024,Zhang2:2024} and the hyperfine splitting due to the crystal electric field gradient has been resolved in precision laser spectroscopy~\cite{Zhang:2024}. This unique isomeric state has inspired numerous proposals for applications~\cite{Tkalya:1996,Tkalya:2011,Tkalya:2022,Peik:2003}, particularly focusing on a nuclear optical clock that could potentially outperform atomic clocks based on electronic excitations~\cite{Peik:2021,Campbell:2012}. The $^{229}$Th nuclear clock would be highly sensitive to new physics because its transition frequency is strongly influenced by both, the nuclear and electromagnetic forces~\cite{Fadeev:2020,Caputo:2024,Beeks:2024b,Banks:2024,Fuchs:2024}.

$^{229}$Th$^{4+}$ doped in a large bandgap solid is a strong candidate for a compact nuclear clock: A far greater number of nuclei can be interrogated in an ionic crystal compared to an ion trap, and the ions in the lattice are deeply in the Lamb-Dicke regime~\cite{Peik:2003}. Optical clocks operate in cycles consisting of state preparation, excitation of the clock transition, and state detection. Ideally, the time used for excitation should dominate the duration of the cycle. %(atoms need to be state-initialized before each interrogation cycle.)
However, in the Th:CaF$_2$ optical clock, the nuclear coherence time of $\approx 10$\,ms, which determines the longest efficient excitation time, is much shorter than the $\approx 600$\,s lifetime of the isomer. This presents an obstacle to the effective operation of a solid-state nuclear clock, %(as the isomer's 10-minute lifetime limits short interrogation cycles,)
thereby deteriorating the expected clock's overall performance~\cite{Kazakov:2012,Nickerson:2021}. Previous studies have demonstrated that quenching, i.e. shortening the lifetime of the isomeric state, is possible using X-rays~\cite{Hiraki:2024}. This is however impractical for the tabletop operation desired in solid-state nuclear clock applications. Furthermore, $^{229}$Th$^{4+}$ ions have a noble gas electronic configuration, which lacks optically accessible electronic transitions that could be used to rapidly quench the isomer nuclear state via an  electronic bridge process~\cite{Porsev:2010}.

A laser-induced quenching (LIQ), as opposed to X-ray-induced quenching, of the isomer state could enable tabletop controlled initialization of the Thorium-229 nucleus into its ground state, which merits further investigation. Aside from radiative decay, the isomeric state can decay via (bound) internal conversion where it transfers its excitation energy to an electron in the surrounding environment, exciting it to (a bound state of) the conduction band, potentially with a much shorter time constant~\cite{Beeks:2021,Nickerson:2020,Nickerson:2021}. Recently, it has been suggested that fluoride vacancies that are created due to irradiating CaF$_2$ can cause electronic levels to emerge, which are resonant with the isomer energy, that can be used for LIQ~\cite{Nalikowski:2024}. 

Here we report the demonstration and extensive investigation of LIQ of the $^{229}$Th isomer population. We show that the isomer state decays unperturbed over a temperature range between 100\,K to 350\,K when no quenching light is present. Furthermore, using LIQ, the isomer can be forced to decay up to three times faster at 297\,K using 30\,mW of average laser power, independent of wavelength between 148\,nm and 420\,nm. At longer wavelengths, at least 729\,nm, LIQ is not present. Finally, LIQ can be suppressed by cooling. LIQ offers an effective and easy-to-implement method for reducing the total duration of the clock interrogation cycle and can improve the stability of a solid-state nuclear clock.

\begin{figure}[h]
\includegraphics[width=0.49\textwidth]{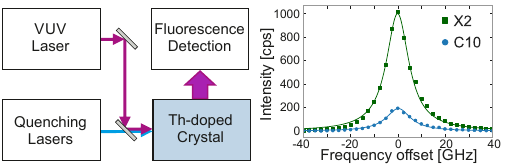}
\caption{\label{fig:spectroscopy} Laser spectroscopy setup and the processed nuclear resonance fluorescence signals of the two $^{229}$Th:CaF$_2$ crystals C10 and X2 with different doping concentration~\cite{Beeks:2022}.}
\end{figure}

The experimental setup consists of a crystal spectroscopy vacuum chamber with a $^{229}$Th-doped crystal and a photomultiplier tube (PMT) fluorescence detection scheme~\cite{Tiedau:2024}, a vacuum ultraviolet (VUV) laser system ~\cite{Thielking:2023}, and a set of lasers emitting at different wavelengths to study laser-induced effects in a Th-doped crystal (see Fig.~\ref{fig:spectroscopy}). Improving on the setup used in~\cite{Tiedau:2024}, we implement a frequency stabilization of the VUV laser at each scanning point. The processed nuclear resonance fluorescence signals of two $^{229}$Th:CaF$_2$ crystals $\left(\mathrm{C10,X2}\right)$ with different doping concentration~\cite{Beeks:2022} are shown in Fig.~\ref{fig:spectroscopy}. The post-processing compensates the asymmetric shape of the recorded spectra when the scanning speed exceeds the decay rate (see~\cite{Tiedau:2024}). Signal records are obtained with 100\,s irradiation and 250\,s detection times.
The C10 crystal was freshly polished and fluorinated which improved its transmission to $>$ 60\,\% at 150\,nm~\cite{Beeks:2023b}. The nuclear transition frequency in CaF$_2$ is measured to be 2020407.8(15)\,GHz which corresponds to an energy of 8.355734(6)\,eV, in agreement with the previous results reported in~\cite{Tiedau:2024,Zhang:2024}. These results also agree with those for $^{229}$Th doped into LiSAF, showing the influence of the host crystal on the nuclear transition frequency to be less than 1~GHz~\cite{UCLA2024}. 

For the investigation of the laser-induced effects in the crystal, we use two laser sources: One is a 10\,ns pulsed tunable Ti:Sa laser with a repetition rate of 1\,kHz and a frequency conversion of its fundamental radiation to the second and third harmonics. This laser is used to irradiate the crystals at 280\,nm, 420\,nm, 729\,nm, and 820\,nm wavelength. The second source is a CW diode laser at 402\,nm. The 402\,nm laser is used to compare between radiation-induced effects caused by a pulsed and a CW laser. The radiation of these lasers superimposed with the VUV radiation and steered to the spectroscopy vacuum chamber. The VUV and quenching light are focused to $\approx 0.5$\,mm diameter at the Th-doped crystal and the overlap of both beams is monitored by a CCD camera. Further details on setup can be found in the Supplemental Material (SM).

Two $^{229}$Th:CaF$_2$ crystals were grown at TU Wien, referred to as ``X2" and ``C10"~\cite{Beeks:2022,Tiedau:2024}: Different pieces of these crystals were used in VUV spectroscopy experiments at SPring-8~\cite{Hiraki:2024}, PTB~\cite{Tiedau:2024} and JILA~\cite{Zhang:2024,Higgins:2024} and demonstrated suitability for precision nuclear laser spectroscopy. The two crystals differ in their $^{229}$Th doping concentration, thus changing the number of nuclei participating in the spectroscopy, as well as the crystal properties. The concentrations are (cm$^{-3}$): $5\cdot10^{18}$ and $3\cdot10^{17}$ with column densities (mm$^{-2}$): $8\cdot10^{15}$ and $1\cdot10^{15}$, for X2 and C10, respectively~\cite{Tiedau:2024}.

\begin{figure}[h]
\includegraphics[width=0.5\textwidth]{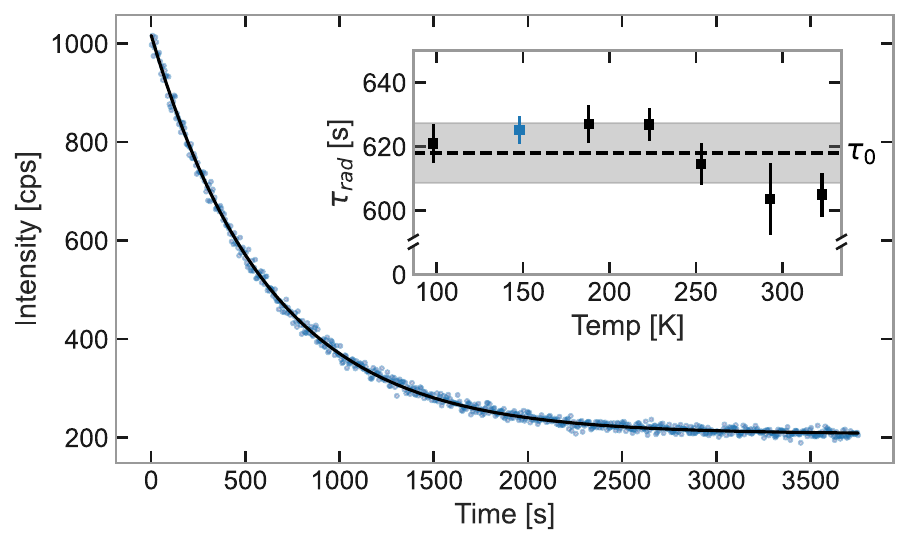}
\caption{\label{fig:decay_temp} $^{229}$Th in CaF$_2$ nuclear fluorescence decay curve after resonant excitation, recorded at 150\,K (C10 crystal). Inset: Radiative isomer lifetime measured at various temperatures. The dashed line corresponds to the mean value and the grey area is the standard deviation.}
\end{figure}

We start by investigating the temperature dependence of the radiative isomer lifetime in CaF$_2$. The C10 crystal, mounted on a cryogenic liquid nitrogen cold finger with a resistive heater, was temperature stabilized at different values between 100\,K and 350\,K. For each temperature, the crystal was irradiated for 500\,s with the on-resonance VUV laser followed by a recording of the isomer fluorescence decay using a PMT. Such a decay curve recorded at 150\,K, as well as an exponential fit, is shown in Fig.~\ref{fig:decay_temp}. No statistically significant changes in the decay constant are observed over the investigated temperature range, as presented in the inset. The average of these lifetime measurements yields $\tau_0=618(9)$\,s which is in agreement with the previously reported lifetime measurements in CaF$_2$~\cite{Tiedau:2024,Hiraki:2024,Zhang:2024}. We observe identical excitation energy and radiative lifetime for the C10 and X2 crystals (further details in the SM).

While we have ascertained that the isomer excitation energy and radiative lifetime do not change with crystal temperature (within experimental uncertainty), we observed a substantial reduction of the isomer signal strength, decreasing by a factor of 3 as the temperature is increased from 100\,K to 350\,K, and laser power and irradiation duration remained constant. This decrease indicates the presence of a competing decay process during excitation that reduces the overall population of excited $^{229}$Th nuclei within the crystal. A similar effect, termed ``X-ray-induced quenching", was observed in~\cite{Hiraki:2024} in which the irradiation of the $^{229}$Th:CaF$_2$ crystal with a 29\,keV X-ray beam reduced the isomer population.

With a spectral power density of $\approx$\,0.1\,$\mathrm{pW/Hz}$, the nuclear laser excitation is performed significantly below the saturation of the isomeric transition. Therefore, while increasing the excitation time $t$ and keeping the VUV laser power constant, the isomer population approaches a steady state level where the excitation rate equals the isomer decay rate. The excitation dynamics can be described by~\cite{Siegman:1986}:
\begin{equation}\label{eq:excitation}
n_{isomer}(t) \approx W \tau \left( 1 - e^{-t/\tau} \right)
\end{equation}
where $n_{isomer}$ is the normalized population of the excited state, $W$ is the excitation rate and $\tau$ the lifetime resulting from the combination of quenching and radiative decay (see SM). 
In the case of pure radiative decay without additional decay channels, $\tau = \tau_0$. Finding an equilibrium timescale $\tau < \tau_0$ with an associated reduction of the fluorescence yield indicates the presence of additional decay channels during the excitation phase, which we refer to as LIQ. 

\begin{figure}[h]
\includegraphics[width=0.5\textwidth]{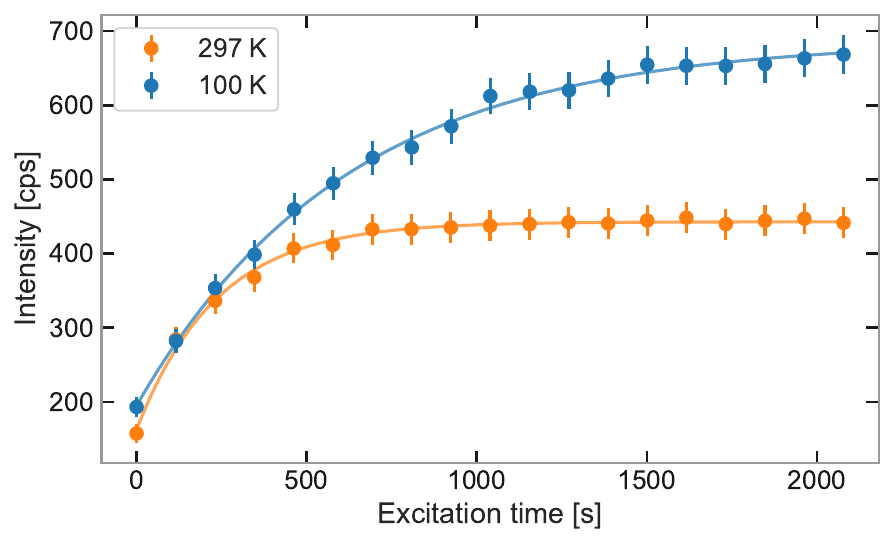}
\caption{\label{fig:excitation_dyn} Excitation dynamics of the $^{229}$Th isomer in the C10 crystal at two different temperatures. Blue: Measured at 100\,K with an extracted $\tau=617(19)$\,s $\approx\tau_0$ and no LIQ observed. Orange: Measured at 297\,K with an extracted $\tau=236(10)$\,s and clear indication of LIQ.}
\end{figure}

We examined the reduction of isomer lifetime of the C10 crystal during excitation at two different temperatures, using eq.~(\ref{eq:excitation}). The results can be seen in Fig.~\ref{fig:excitation_dyn}. 
At 100\,K the time constant $\tau$ matches the radiative decay constant $\tau_0$, indicating no LIQ. At room temperature, the time constant $\tau$ was reduced by a factor of 2.6, indicating an additional decay process present during laser irradiation. 

We conjectured that the LIQ observed in the excitation dynamics is related to the intense off-resonant VUV laser radiation. To study LIQ independently from the excitation dynamics, we implemented isomer quenching after nuclear excitation, primarily using the higher-doped X2 crystal. C10 behaves qualitatively the same as X2, however, there is a significant quantitative difference with stronger LIQ observed for the lower-doped crystal.

For these measurements we excited the isomer on resonance for up to 500\,s, followed by fluorescence detection over a period of 60\,s. Thereafter, the isomer was exposed to off-resonant laser light for up to 200\,s where the VUV center frequency was shifted by approximately 100\,GHz. Additionally, we studied LIQ using wavelengths between 280 and 840\,nm. The off-resonance light does not induce further excitation of the isomer, but would cause LIQ. 

\begin{figure}[h]
\includegraphics[width=0.5\textwidth]{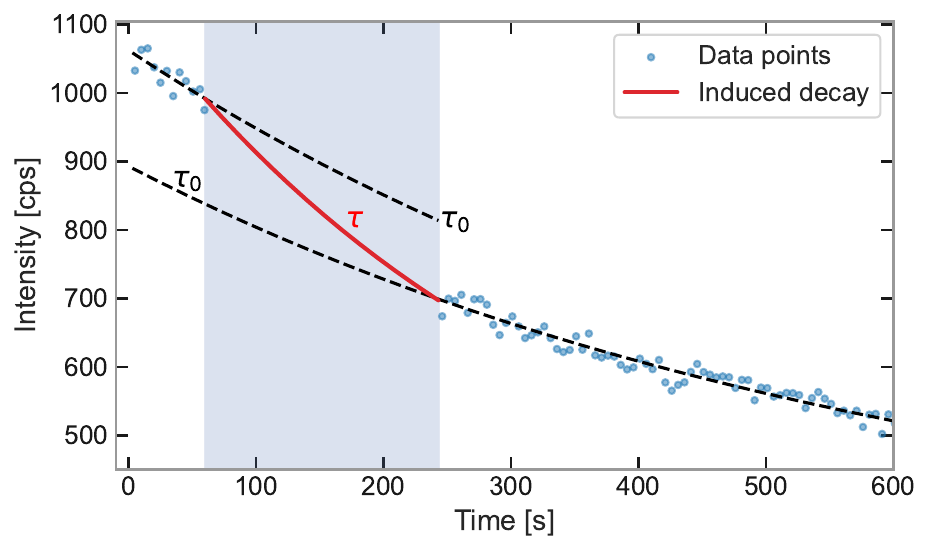}
\caption{\label{fig:off_res} Quenching of the $^{229}$Th isomer with laser radiation: Exponential isomer decay signal (X2 crystal) with the radiative lifetime $\tau_0$ recorded before and after off-resonant laser irradiation. Blue shaded area: Time window in which the crystal was irradiated off-resonantly. Red curve: Laser-induced exponential decay with a lifetime $\tau<\tau_0$.}
\end{figure}

A typical result obtained with the off-resonance scheme is depicted in Figure~\ref{fig:off_res}; the dashed black lines indicate the exponential decay of the isomer, characterized by the radiative lifetime $\tau_0$, both before and after the off-resonant laser irradiation. In the blue shaded area, during the off-resonance laser irradiation, an isomer lifetime of $\tau < \tau_0$ can be inferred, as shown by the red curve. No fluorescence signal could be taken during the off-resonant laser irradiation period because of stray light blinding the PMT.

Utilizing the above mentioned off-resonance scheme, we investigated the LIQ in the X2 crystal for different temperatures and with two different lasers: the off-resonant pulsed VUV excitation laser ~\cite{Tiedau:2024,Thielking:2023} operating at an average power of 0.2\,mW, and a CW laser with a wavelength of 402\,nm and a power of 10\,mW. For the C10 crystal, the temperature dependence was solely investigated using the VUV excitation laser. We scanned a temperature range from 170\,K to 340\,K and extracted the reduced lifetime $\tau$.

\begin{figure}[h]
\includegraphics[width=0.5\textwidth]{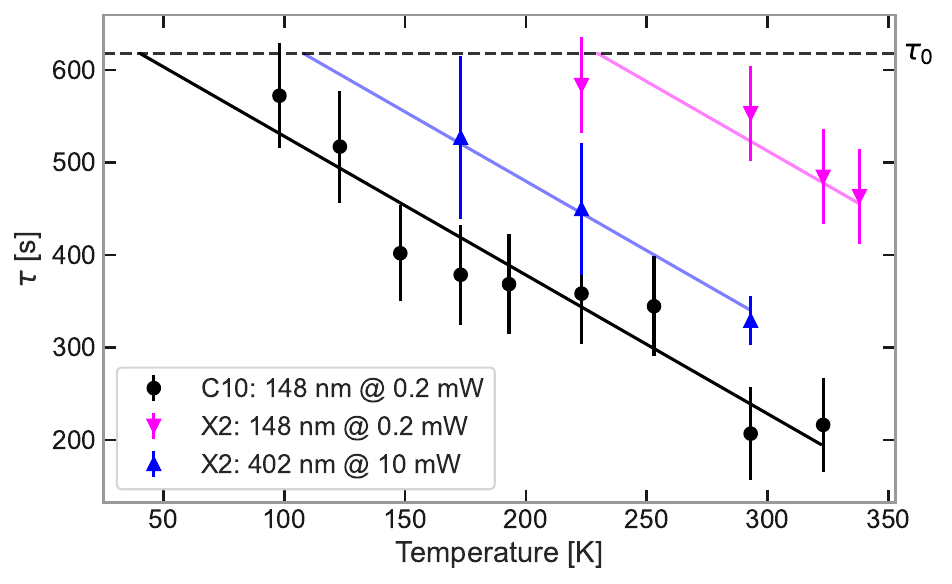}
\caption{\label{fig:tau_T} Isomer lifetime $\tau$ affected by laser quenching for different temperatures (X2 and C10 crystal). Dashed black line: Unperturbed radiative lifetime $\tau_0$. Blue: LIQ was induced by a 402\,nm CW laser with a power of 10\,mW. Purple: LIQ was induced using the pulsed VUV laser shifted by 100\,GHz away from the isomer transition frequency. Linear fits to guide the eye have imposed identical slope, yielding -1.5\,s/K.}
\end{figure}

A clear temperature dependence of $\tau$ was observed, as illustrated in Fig.~\ref{fig:tau_T}. Regardless of laser power, the observed temperature dependence appeared consistent, with stronger LIQ observed at higher temperatures, while LIQ was significantly reduced or completely suppressed at lower temperatures. The quantitative difference in the LIQ between the two types of lasers can be attributed to the large power difference between both lasers, which is investigated further below. The quenching rate difference between X2 and C10 is subject to further investigation. Note that in the case of the off-resonant 148\,nm VUV laser, the quenching light is intrinsically aligned to the isomeric nuclei produced inside the crystal while in the case of the 402\,nm laser, the alignment is non-trivial, possibly underestimating the LIQ effect for a given laser power. At the lowest investigated temperature of 100\,K, a full suppression of the LIQ could be observed for all investigated laser sources and powers.

\begin{figure}[h]
\includegraphics[width=0.5\textwidth]{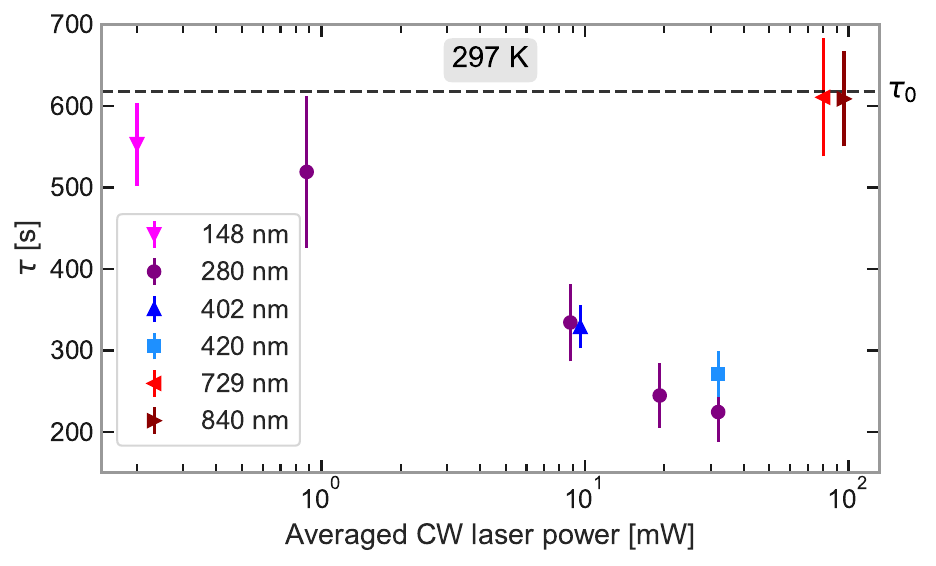}
\caption{\label{fig:tau_LP} Off-resonant LIQ in the X2 crystal at different laser powers and wavelengths, at room temperature. The 402\,nm laser is CW while all other lasers are pulsed. The dashed black line shows the unperturbed radiative lifetime $\tau_0$.}
\end{figure}

To investigate the power dependence of LIQ, we employed the same off-resonance scheme while maintaining a constant temperature of approximately 297\,K and varying the laser power. This investigation utilized five different laser wavelengths: The pulsed VUV laser detuned by 100\,GHz from the isomer transition with a 0.2\,mW average power, a 280\,nm pulsed laser with adjustable power, a 402\,nm CW laser with 10\,mW, a 420\,nm pulsed laser with 32\,mW, and an infrared laser operating at 729\,nm and 840\,nm with 80\,mW and 96\,mW, respectively. 
It is important to note that, with the exception of the VUV laser, the alignment for each of these lasers in the crystal varies and the effective laser power can be reduced by up to 50\,\%, caused by imperfect overlap of the quenching laser beam with the excited nuclei.
Therefore, it is impossible to give a precise quantitative comparison between different lasers. Figure~\ref{fig:tau_LP} presents the isomer lifetime $\tau$ as a function of laser power for different laser wavelengths. For wavelengths between 148\,nm and 420\,nm the power dependence of LIQ appears to be independent of wavelength. At near infrared wavelengths of 729\,nm and above, even a 100\,mW laser fails to induce LIQ. The LIQ process seems to be activated above a (yet to be precisely determined) photon energy threshold in the visible range.   

%A clear power dependence of LIQ, which appears to be independent of wavelength up to a (yet to be precisely determined) threshold, can be observed. Beyond the wavelength threshold, even a 100\,mW infrared laser fails to cause LIQ.

\begin{figure}[h]
\includegraphics[width=0.5\textwidth]{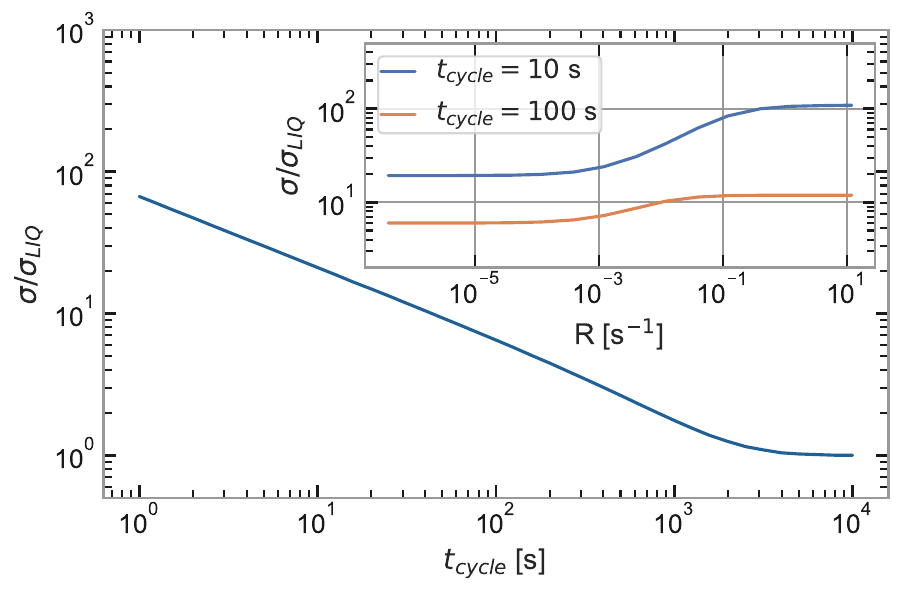}
\caption{\label{fig:sigma}
Ratio of fractional frequency instabilities of a $^{229}$Th:CaF$_2$ clock without ($\sigma$) and with LIQ ($\sigma_{LIQ}$) based on Ref.~\cite{Kazakov:2012} as a function of a single interrogation cycle time for an excitation rate $R=\gamma$.
Inset: The same ratio but as a function of excitation rate R for $10\,$s and $100\,$s as cycle time ($t_{cycle}$).}
\end{figure}

The performance of a $^{229}$Th solid-state clock was estimated using a fluorescence-based laser interrogation scheme in Ref.~\cite{Kazakov:2012}. In this scheme, the optimal short-term stability is attained when the time of a single clock cycle is significantly longer than the radiative lifetime $\tau_0$. It is necessary to transfer the excited nuclei back to the ground state. This makes clock operation strongly dependent on the intrinsic stability of a laser used for excitation~\cite{Santarelli98}. 
However, LIQ offers the option to depump the isomer population to the ground state, which allows to significantly reduce the clock cycle. 
Figure~\ref{fig:sigma} shows the ratio of the fractional frequency instability with and without LIQ for instantaneous quenching.

Quenching operation allows us to reach lower fractional frequency instability at shorter cycle times. This significantly reduces the operation time, which will relax the constraints on the experimental stability. This is particularly important in searching for new physics \cite{safronova2018search} in the high-frequency domain, where a combination of a short probing time and a high sensitivity is crucial.

To conclude, we have demonstrated that a broad range of laser wavelengths can induce accelerated decay of the $^{229}$Th isomer. %This LIQ is caused by interactions of the nucleus with the perturbed CaF$_2$ solid-state host. 
Within our experimental conditions, LIQ is largely independent of wavelength below 420\,nm, and is completely absent at wavelengths of 729\,nm and above. We observe an approximately linear scaling of LIQ with laser power. Additionally, LIQ is thermally activated with an activation temperature above 100\,K.

The underlying physical process of LIQ is currently unknown and will be the focus of further investigations, in particular for different host crystals. Our study shows that it requires two factors, possibly proceeding in two separate steps: a (single) photon of approximately 2\,eV energy and thermal activation in the range of 0.02\,eV. The latter might provide mobility to (laser-induced) defect centers~\cite{Rix:2011,Nalikowski:2024} within the crystal. These might create intragap electronic states that can interact with the nucleus via electron bridge processes~\cite{Nickerson:2020,Nickerson:2021}. Another possibility is the creation of phonons which mediate the internal conversion process by ensuring conservation of energy and momentum.

%Additionally, LIQ is highly temperature-sensitive, with full suppression achievable at cryogenic temperatures. At the current state, the exact interaction of excited nuclei with the electronic environment that causes the LIQ, possibly through an electron bridge process~\cite{Nickerson:2020,Nickerson:2021}, is unknown, and needs to be further investigated. A possible cause is the laser-induced creation of defect centers~\cite{Rix:2011,Nalikowski:2024}, which create intragap electronic excited states at or close to the nuclear excitation energy to which the nucleus can (resonantly) couple to. Another possibility is the creation of phonons which mediate the internal conversion process by ensuring conservation of energy and momentum.

Regardless of the specific cause, LIQ of the excited $^{229}$Th isomer presents several intriguing applications. Firstly, it demonstrates that the excited nucleus can interact with the crystal environment and potential defect states, offering a different and possibly more efficient excitation pathway. Secondly, LIQ opens up the possibility of initializing the $^{229}$Th nucleus into its ground state in a $^{229}$Th:CaF$_2$ crystal and with that, an enhanced clock performance~\cite{Kazakov:2012}. Additionally, LIQ could provide an explanation for previous unsuccessful attempts to excite and detect the isomer using broadband VUV synchrotron sources~\cite{Stellmer:2018,Jeet:2015}: The significant off-resonance fraction of photons together with the high crystal temperature probably induced significant LIQ. Future solid-state-based $^{229}$Th nuclear excitation approaches invoking broadband radiation or multi-photon schemes will require cooling of the crystal and/or suppression of non-resonant photons. \\

%Repeating the experiment with the crystal at 100\,K would reduce LIQ and potentially provide a signal. 

%These findings introduce a toolbox for the control of the $^{229}$Th isomer lifetime through laser irradiation and temperature, which is crucial for the operation of a solid-state nuclear clock.

\begin{acknowledgments}
We would like to thank J. Thielking for support in early phases of the experiment, N. Huntemann for providing an Yb-clock frequency reference, and T. Leder and M. Menzel for the design and manufacture of the mechanical structures.
%T. Leder and M. Menzel for the design and manufacture of the mechanical structures, N. Huntemann for providing an Yb-clock frequency reference for the wavelength measurements, 
We further thank A. Leitner, N. Hosseini, J. Sterba, M. Pressler, V. Rosecker, D. Hainz and M. Veit-\"Oller for support in the preparation, characterization, and handling of radioactive samples. 

This work has been funded by the European Research Council (ERC) under the European Union’s Horizon 2020 research and innovation programme (Grant Agreement No. 856415) and the Austrian Science Fund (FWF) [Grant DOI: 10.55776/F1004, 10.55776/J4834, 10.55776/PIN9526523]. The project 23FUN03 HIOC [Grant DOI: 10.13039/100019599] has received funding from the European Partnership on Metrology, co-financed from the European Union’s Horizon Europe Research and Innovation Program and by the Participating States. We acknowledge support by the Deutsche Forschungsgemeinschaft (DFG) – SFB 1227 - Project-ID 274200144 (Project B04), and by the Max-Planck-RIKEN-PTB-Center for Time, Constants and Fundamental Symmetries. K.B. acknowledges support from the Schweizerischer Nationalfonds (SNF), fund 514788 “Wavefunction engineering for controlled nuclear decays."

\end{acknowledgments}

%\bibliographystyle{apsrev4-1}
%\bibliography{bibliography.bib}

%

\end{document}

\flushbottom
\title{Supplemental Material:}

\author{F. Schaden}
\author{T. Riebner}
\author{I. Morawetz}
\author{L. Toscani De Col}
\author{G.A. Kazakov}
\author{K. Beeks}
\author{T. Sikorsky}
\author{T. Schumm}
\email{email: thorsten.schumm@tuwien.ac.at}
\affiliation{Vienna Center for Quantum
Science and Technology, Atominstitut, TU Wien, 1020 Vienna, Austria.}

 \author{K. Zhang}
\author{V. Lal}
\author{G. Zitzer}
\author{J. Tiedau}
\author{M.V. Okhapkin}
\author{E. Peik}
\email{email: ekkehard.peik@ptb.de}

\affiliation{Physikalisch-Technische Bundesanstalt, 38116 Braunschweig, Germany}

\date{\today}
\maketitle

\subsection{Experimental setup}
The laser source for the nuclear spectroscopy uses four-wave mixing in a xenon gas cell for VUV production~\cite{Thielking:2023}. The two seed lasers used are two CW Ti:sapphire (Ti:Sa) ring lasers in combination with two pulsed dye amplifiers and a third harmonic generation (THG) unit.

The frequency of the first CW Ti:sapphire laser is stabilized to excite, after frequency tripling (frequency $\Omega_1$), the two-photon resonance of Xe at 80119\,cm$^{-1}$. The second CW Ti:sapphire laser (operating frequency $\Omega_2$) is tunable to define the difference frequency of the four-wave mixing process. The two dye amplifiers operate at a repetition rate of 30\,Hz and a pulse duration of $\approx$\,10\,ns.

Both beams at frequencies $\Omega_{1}$ and $\Omega_{2}$ are focused into the Xe cell using an achromatic objective. The VUV radiation frequency of the four-wave mixing process is defined by $\Omega_{VUV}=2\times\Omega_1-\Omega_2$ (see~\cite{Thielking:2023}).
The mean pulse energy of the source is $\approx$\,10\,$\mu$J for the wavelength of 148.4\,nm which corresponds to the nuclear transition in $^{229}$Th~\cite{Tiedau:2024} and the measured spectral linewidth of the VUV source is $\leq$\,10\,GHz. This corresponds to a VUV power spectral density of $\approx$\,0.1\,$\mathrm{pW/Hz}$. The VUV radiation is focused to $\approx$\,0.5\,mm diameter in the thorium doped crystal by a MgF$_{2}$ lens.

For the experiments we use Th-doped crystals with two different concentrations as described in Ref.~\cite{Tiedau:2024}. Here we use the same crystal codes, where C10 and X2 crystals have a concentration of $3\times10^{17}$cm$^{-3}$ and $5\times10^{18}$cm$^{-3}$, respectively.
The C10 crystal is freshly fluorinated and repolished ~\cite{Beeks:2023b} to recover its transmission after damaging~\cite{Beeks:2022} through long VUV exposure times during the search of the isomer energy ~\cite{Tiedau:2024}. The improved transmission of C10 is $\approx 60\,\%$ at 150\,nm. Previously~\cite{Tiedau:2024}, the signal ratio between X2 and C10 was a factor of 25, now it is 5. Considering that the ratio of the column densities is 8, we conclude that the previously observed discrepancy is now resolved and was due to crystal imperfections which were fixed by fluorination and polishing~\cite{Beeks:2024}.

The Th-doped crystal is attached to a liquid nitrogen cooled cold finger that is mounted on a 3-axes vacuum manipulator. In addition, a resistive heater is attached to the crystal holder to control the crystal temperature for measurements of the isomer decay in a wide temperature range (100-350\,K).
As in the previous experiment, A spherical mirror combined with two dielectric mirrors and a short focal length MgF$_2$ lens are used to filter, direct and focus the signal onto a Cs-I photomultiplier tube (PMT). Additionally, we use a lead shield to protect the PMT from direct exposure of $\gamma$-radiation produced by the crystal. We use a CCD camera to observe photo- and radioluminescence for monitoring the crystal and positioning the laser beam.

\subsection{Isomer resonance linewidth}

We repeat the resonant excitation of the Th-229 nucleus of the two differently doped crystals, C10 and X2, with an improved stabilization of the irradiation laser. A frequency stabilization of the difference frequency laser ($\Omega_2$) at each scanning point over a scanning range of 90\,GHz was implemented. This allows us to control the exact frequency of the vacuum ultraviolet (VUV) spectroscopy laser. 

The crystal was irradiated for $t_{e}=100$\,s at each frequency step, followed by an acquisition period of $t_{d}=250$\,s.
As the isomer has a long decay time of more than 600\,s in the CaF$_2$ host, and the scanning speed is significantly shorter, this measurement procedure leads to an asymmetric shape of the recorded spectra~\cite{Tiedau:2024}. This effect can be compensated in post-processing by removing background counts and isomer decay from previous points.
%The processed data $y_\mathrm{i}$ for both X2 and C10 obtained with scan steps of 2.5\,GHz of the VUV laser are shown \red{in the main text.} 

We implemented a set of VUV filters to investigate any potential fluorescence between 150\,nm and 200\,nm, as was observed in Ref.~\cite{UCLA2024,Kraemer:2022,Beeks:2022}. While exciting the isomer at 148.4\,nm, no fluorescence signal above the PMT dark noise level of 2.5(5) counts per second was detected. %Therefore, the fluorinated crystal contains fewer defects than previous samples.

\subsection{Isomer photon yield as a function of temperature}

\begin{figure}[h]
\includegraphics[width=0.5\textwidth]{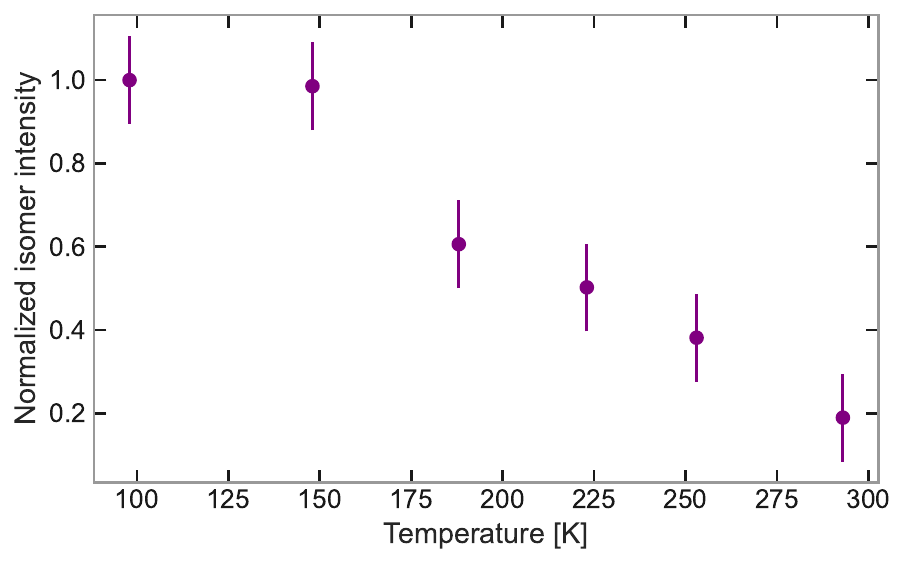}
\caption{\label{fig:amp_vs_T} Isomer fluorescence intensity normalized as a function of crystal temperature for C10.}
\end{figure}

We measured the decay curve of the isomer doped in the C10 crystal over various temperatures. While the fluorescence lifetime remained constant, we observed a significant decrease in isomer fluorescence intensity. As the temperature increases, the intensity declines. At low temperatures, the intensity saturates as it is shown in Fig.~\ref{fig:amp_vs_T}. This strongly suggests the presence of a competing decay process of the isomer during laser excitation.

%Show plot of A vs T for C10 and mention competing process
\subsection{Rate equations for excitation process}

To investigate the excitation dynamics and a possible competing decay process during laser excitation, we solve the rate equations for a 2-level system (see for example~\cite{Siegman:1986}).

%\begin{gather}
%\frac{d n_2(t)}{dt} = W \cdot n_1(t) - W \cdot n_2(t) - \gamma \cdot n_2(t), \\
%n_1(t) + n_2(t) = 1
%\end{gather}

%\begin{align}
%\frac{d n_2(t)}{dt} &= W \cdot n_1(t) - W \cdot n_2(t) - \gamma \cdot n_2(t), \\
%n_1(t) + n_2(t) &= 1
%\end{align}

%where $\gamma=\gamma_{id}+\gamma_{ic}$, $\gamma_{id}$, $\gamma_{ic}$ are the isomer decay rate and the internal conversion decay rate respectively, $n_1$ and $n_2$ are the populations of the ground and the excited states respectively and $W$ is the excitation rate. 
Solving this differential equation gives the time dependent normalized population of the excited state:
\begin{equation}\label{eq:sol_n2}
{n_2}(t)=\frac{W(1-e^{-(2W+\gamma)t})}{2W+\gamma}
\end{equation}

where $\gamma=\gamma_{0}+\gamma_{q}= 1/\tau$, $\gamma_{0}= 1/\tau_0$, $\gamma_{q}$ are the isomer decay rate and the quenching decay rate, respectively, $n_2$ is the population of the excited state and $W$ is the excitation rate. 

The excitation rate per ion without quenching is given by:
\begin{equation}
W=\frac{2\pi^3c^2\gamma_{0}}{h\omega^3}\frac{P_p\tau_{p}f_{rep}}{S\Delta_\nu}\sim2\times10^{-8}\mathrm{s}^{-1},
\end{equation}
where $P_p$ is the VUV pulse power, $\tau_p$ is the VUV pulse duration, $f_{rep}$ is the VUV pulse repetition rate, $S$ is the beam area and $\Delta_\nu$ is the laser linewidth.

Since the excitation rate in our experiment is significantly smaller than the spontaneous radiative decay of the nuclear isomer, $W\ll\gamma$, equation (\ref{eq:sol_n2}) can be simplified to 
\begin{equation}\label{eq:sol}
{n_2}(t)\approx\frac{W(1-e^{-\gamma t})}{\gamma}=W\tau(1-e^{-t/\tau})
\end{equation}

Therefore, while increasing the excitation time and keeping the VUV power constant, the isomer fluorescence signal approaches a steady state level ($W \tau$) for which the excitation rate equals the isomer decay rate, in the absence of a competing decay channel. 
\\

\bibliographystyle{apsrev4-1}
\bibliography{bibliography.bib}